\documentclass[prx,superscriptaddress,aps, twocolumn, reprint, showpacs, longbibliography, footnoography]{revtex4-2}
\usepackage{etex}

\usepackage{amsmath, amssymb, amsfonts, braket, commath, bm, empheq, array} 

\usepackage[]{graphicx}
\usepackage{hyperref}
\hypersetup{
	colorlinks=true,
	linkcolor=red,
	filecolor=blue,
	urlcolor=magenta,
}

\usepackage{dblfloatfix}

\newcommand{\bte}{\ensuremath{\beta\textrm{ ensemble}}}

\def\eg{\ensuremath E_{\textrm{G}}}			

\newcommand{\gat}{\ensuremath{\gamma_{\text{AT}}}}	
\newcommand{\get}{\ensuremath{\gamma_{\text{ET}}}}	
\newcommand{\Hm}{ \ensuremath{\overline{H}} }		
\newcommand{\ipr}{\ensuremath{\textrm{I}}}
\newcommand{\iloc}{\ensuremath{i_\textrm{loc}}}

\newcommand\lrp[1]{\left(#1\right)}

\newcommand{\lra}{\;\; \Leftrightarrow \; \;}
\newcommand{\mean}[1]{\ensuremath{\left\langle#1\right\rangle}}
\newcommand{\prob}[1]{\ensuremath{\text{P}\del{#1}}}

\begin{document}
\title{Robust non-ergodicity of ground state in the \bte}
\author{Adway Kumar Das}\email{akd19rs062@iiserkol.ac.in}
\author{Anandamohan Ghosh}\email{anandamohan@iiserkol.ac.in}
\affiliation{
	Indian Institute of Science Education and Research Kolkata, Mohanpur, 741246 India
}
\author{Ivan M. Khaymovich}\email{ivan.khaymovich@gmail.com}
\affiliation{Nordita, Stockholm University and KTH Royal Institute of Technology Hannes Alfv\'ens v\"ag 12, SE-106 91 Stockholm, Sweden}
\affiliation{Institute for Physics of Microstructures, Russian Academy of Sciences, 603950 Nizhny Novgorod, GSP-105, Russia}

\date{\today}
\begin{abstract}
In various chaotic quantum many-body systems, the ground states show non-trivial athermal behavior despite the bulk states exhibiting thermalization. Such athermal states play a crucial role in quantum information theory and its applications. Moreover, any generic quantum many-body system in the Krylov basis is represented by a tridiagonal Lanczos Hamiltonian, which is analogous to the matrices from the \bte, a well-studied random matrix model with level repulsion tunable via the parameter $\beta$. Motivated by this, here we focus on the localization properties of the ground and anti-ground states of the \bte. Both analytically and numerically, we show that both the edge states demonstrate non-ergodic (fractal) properties for $\beta\sim\mathcal{O}(1)$ while the typical bulk states are ergodic. Surprisingly, the fractal dimension of the edge states remain three time smaller than that of the bulk states irrespective of the global phase of the \bte. 
In addition to the fractal dimensions, we also consider the distribution of the localization centers of the spectral edge states, their mutual separation, as well as the spatial and correlation properties of the first excited states.
\end{abstract}
\maketitle
\section{Introduction}\label{sec_intro}
The physics of thermalization in isolated quantum many-body systems has intrigued the condensed matter community over the last few decades~\cite{Deutsch1991,Srednicki1994,Srednicki1996,rigol2008thermalization}. The quantum analogue of the classical Boltzmann ergodicity hypothesis, named as the eigenstate thermalization hypothesis (ETH) claims that in general, an isolated quantum chaotic system will locally thermalize under its own unitary evolution where any local information on the initial state will be lost. Such generic systems play a crucial role in quantum information theory due to their non-equilibrium dynamics~\cite{Polkovnikov_2011}.

Typically ETH is probed for the highly excited bulk states lying in the center of the energy spectrum where the density of states has the largest value. However, the ground states are particularly important as they play a major role in the conventional (low-energy) physics of complex and correlated systems while driving collective phenomena like superconductivity~\cite{Bardeen1957}, fractional quantum Hall effect~\cite{Laughlin1983} and many others. Compared to the bulk states, the ground states are also more accessible analytically, e.g.~through renormalization group methods~\cite{Monthus2015}, numerically using tensor-network and quantum Monte Carlo methods~\cite{Luitz2014} and experimentally in quantum simulators using variational algorithms~\cite{Moll2018} in the current noisy intermediate-scale quantum era. In this sense, some counter-intuitive properties of the ground states of generic many-body systems bring a lot of attention since a series of seminal works~\cite{Stephan2009, AtasBogomolny2012_MF_GS, AtasBogomolny_MF_GS_detailed}.
Indeed, in case of paradigmatic models like Ising and Heisenberg spin-$\frac{1}{2}$ chains~\cite{Vidal2003, AtasBogomolny2012_MF_GS, Monthus2015}, Bose-Hubbard model~\cite{Lindinger2019}, the ground states do not show ergodic behavior despite the mid-spectrum states exhibiting thermalization. Instead, the ground states demonstrate ergodicity breaking both in terms of the equipartition over degrees of freedom (i.e.~being fractal states that occupy only measure zero of all the Hilbert-space configurations) and entanglement~\cite{Amico2008}.  The ground state entanglement leads to long-range correlations at zero temperature producing quantum phase transition e.g.~in Ising~\cite{Osterloh2002, Osborne2002}, Lipkin-Meshkov-Glick model~\cite{Wu2004} and Mott insulator-superfluid transitions in bosonic systems~\cite{Sachdev1999book}. In addition, the empirical probability distribution of ground states for various disordered many-body systems has been recently addressed in several papers, see, e.g.~\cite{Buijsman3}.

The many-body systems are notorious to handle for larger system sizes due to exponentially increasing Hilbert space dimension. Followed by a recent analogy between a tridiagonal Lanczos Hamiltonian in the basis of Krylov operators for a generic quantum many-body system and the \bte~\cite{Bala2022,Parker2019universal}, we consider the ground-state properties of the later, being the proxy of the above many-body interacting systems. In case of the \bte, the ground state energy can be related to the stochastic Airy operators~\cite{Ramirez1}, while its density can be expressed in terms of the multivariate integrals~\cite{Albrecht1}. The relevant Tracy-Widom laws for arbitrary $\beta$ have been extensively studied~\cite{Dumaz1, Borot1, Allez1, Edelman4, Forrester9}. 
Particularly, for $N\to \infty$ and $\beta\to 0$, the Tracy-Widom distributions weakly converge to the Gumbel distribution~\cite{Johansson2, Forrester6}. Recently a recursion relation is proposed for the probability distribution of the ground state energy of the Laguerre \bte~\cite{Kumar2}.

In this work, we look at the structure of the ground state of the Gaussian \bte\ and demonstrate its non-ergodic structure even in the case of $\beta \sim \mathcal{O}(1)$, when the typical bulk states are known to be ergodic~\cite{Das5}.

The rest of the paper is organized as follows.
Section~\ref{sec_model} describes the model in focus, summarizing the main properties of the \bte. In Sec.~\ref{sec_GS_NEE}, we consider both numerically and analytically the ground-state properties of the non-ergodic phase of the model, focusing on the fractal dimensions of the ground state and the statistics of the localization centers.
In Sec.~\ref{sec_Erg}, we look at the ground state in the extended phase ($\beta\gtrsim 1$), shown to be dominated by a deterministic part of the Hamiltonian.
Section~\ref{sec_concl} summarizes our results.

\section{Model}\label{sec_model}
\subsection{Model description}
The \bte\ is a random matrix model with a joint probability distribution of eigenvalues known exactly and controlled only by the Dyson's index $\beta$, interpreted  as the inverse temperature of an equivalent Coulomb gas model~\cite{Dyson2, Forrester3}. Consequently, the standard joint probability distributions of the Gaussian ensembles, with $\beta=1$, $2$, and $4$, depending on the symmetry, are generalized over any real $\beta$-values in case of the \bte. Such an ensemble consists of real symmetric tridiagonal matrices having independent random elements~\cite{Dumitriu1}. Corresponding symmetric $N\times N$ Hamiltonian matrix $H$ has the following non-zero elements, $H_{m,n}$, $1\leq m\leq n\leq N$,
\begin{align}
	\label{eq_H_def}
	H_{n,n} \sim \mathcal{N}(0, 1),\quad H_{n, n+1} 
= y_n,\: \sqrt{2}y_n \sim \chi_{n\beta} \ ,
\end{align}
where the diagonal elements obey the Gaussian distribution $\mathcal{N}(0, 1)$, with zero mean and unit variance, while the off-diagonal ones follow the Chi distribution $\chi_{n\beta}$, with the site index $n$-dependent width parameter $n\beta$, $1\leq n\leq N$. 
In the above site basis, $H$ represents a random single-particle model on a $1$d lattice with open boundary conditions and site-dependent hopping term. The relative strength of the on-site ($H_{n,n}$) and the hopping terms ($y_n$) at a typical site $n\sim \mathcal{O}(N)$ indicate a suitable scaling of the system parameter $\beta = N^{-\gamma}$, where $\gamma$ is a certain real-valued parameter~\cite{Das2, Das5, nakano2018gaussian}.

\subsection{Bulk phase diagram}
Equation~\eqref{eq_H_def} implies that the hopping amplitudes increase on average along the lattice as $\mean{y_n} = \sqrt{\beta n}$ for $\beta n \gtrsim 1$ and present a highly inhomogeneous system. Such inhomogeneity allows for phase transitions in the \bte, whereas criticality is forbidden in a generic $1$d system with uncorrelated short-range hopping~\cite{Cuesta1, Albert1}. Particularly, for all the typical bulk states, there are the Anderson transition at $\gamma = 1$ and the ergodicity breaking transition at $\gamma = 0$, leading to non-ergodic extended (NEE) phase for $0 < \gamma < 1$. 
Therefore, other than long-range hopping \cite{Kravtsov1, Deng2}, quasiperiodicity \cite{Evangelou1}, drive \cite{Ray1}, interaction \cite{Torres2} or correlated disorder \cite{Nosov1}, inhomogeneity in short-range hopping strength is another ingredient for phase transition in 1d systems. Apart from the hopping terms, inhomogeneous interaction~\cite{wang2023arxiv} or inhomogeneity in on-site terms can lead to criticality as well, e.g.~Wannier-Stark localization induced by linear potential~\cite{Nieuwenburg1}.

In the ergodic regime of the \bte, the bulk eigenstates span over the entire Hilbert space and the eigenvalues are correlated over distances much larger than the mean level spacing~\cite{Bohigas1}. The degree of energy correlation is controlled by $\beta$~\cite{Mehta1} such that the standard Wigner-Dyson level repulsion is observed for $\beta\sim\mathcal{O}(1)$. Contrarily, the localized phase ($\gamma>1$) reflects the characteristics of the integrable systems where all the eigenvalues are uncorrelated and the eigenstates are spatially localized~\cite{Berry3}.

In the NEE phase, the typical bulk eigenstates are fractal with the dimension $D_2=1-\gamma$, given by the scaling of the inverse participation ratio (IPR), which is the 2nd moment of the density of the eigenstate intensities
\begin{gather}\label{eq_IPRq}
	\ipr_2 = \sum_n |\Psi_E(n)|^{4}\sim N^{-D_2}
\end{gather}
where $\Psi_E(n)$ is the $n$th component of the eigenstate at energy $E$ in the chosen basis.
The fractional value $0<D_2<1$ implies that the typical bulk states occupy an extensive part but vanishing fraction of the Hilbert space. Such NEE states are observed in various
random matrix ensembles~\cite{Kravtsov1,Biroli_RP,Ossipov_EPL2016_H+V, Monthus, vonSoosten2017non, BogomolnyRP2018,
Venturelli1, BirTar_Levy-RP, Buijsman2022circular, Khaymovich1, Das1, Das3, Das4, Evers1, Mirlin1, Cizeau1, Nosov1, Khaymovich2, Tang1, Roy3},
physical models~\cite{Luca1, Kravtsov3, Pino2, Altshuler3, Laumann2014QREM, Kutlin1, Mace1, Luitz1, Luitz3, Torres2, Ray1, Wang3, Roy2, Motamarri1,Cai2013AA+p-wave,DeGottardi2013AA+p-wave,Wang2016AA+p-wave,Fraxanet2021AA+p-wave,Fraxanet2022AA+p-wave} and realized in experimental setups~\cite{Gao1, Xu1,Dietz2023RP_exp}.
However, in the NEE phase of the \bte, nearby eigenvalues remain uncorrelated with non-hybridizing eigenstates but two distant eigenvalues separated by $\Delta E > N^{\frac{\gamma - 1}{2}}$ can be correlated~\cite{Relano3}. Moreover, in the middle of the spectrum, a small energy window, $(-\eg, \eg)$, $\eg\sim \mathcal{O}(1)$, contains $\mathcal{O}(N^\gamma)$ localized states along with the NEE states without forming any mobility edge~\cite{Das5}, therefore challenging Mott's argument~\cite{Mott1}. Such emergence of localized or extended states at a given energy in different realizations of \bte\ for $0<\gamma<1$ can be attributed to three ingredients: residual level repulsion, non-ergodicity of most bulk states, and spatial separation of extended and localized states. These criteria have been recently identified and demonstrated in case of the \bte~\cite{Das5}. The distinct eigenstate and eigenvalue structure in the NEE phase of \bte\ can be analytically understood from a spatially local mapping to the $1$d Anderson model with $N$-dependent hopping strength~\cite{Das5}. Such spatially local mapping can also explain the behavior of the ground state of \bte, as we will show in this work.

\section{Ground state in the NEE phase}\label{sec_GS_NEE}
We start this section with the numerical exploration of the ground $\ket{\Psi^-}$ and anti-ground $\ket{\Psi^+}$  eigenstates of~\eqref{eq_H_def} at both the smallest and the largest eigenenergies. In Subsecs.~\ref{subsec_GS_Dq} and~\ref{subsec_GS_iloc}, we focus on the scaling properties, the spatial spread and the localization-center distribution, of $\ket{\Psi^\pm}$ separately. Subsection~\ref{subsec_GS_overlap} is devoted to the correlations between $\ket{\Psi^+}$ and $\ket{\Psi^-}$.
In the  Subsec.~\ref{subsec_GS_analytics}, we provide analytical explanation of the numerical results.

\begin{figure}[t]
	\centering
	\includegraphics[width=0.99\columnwidth]{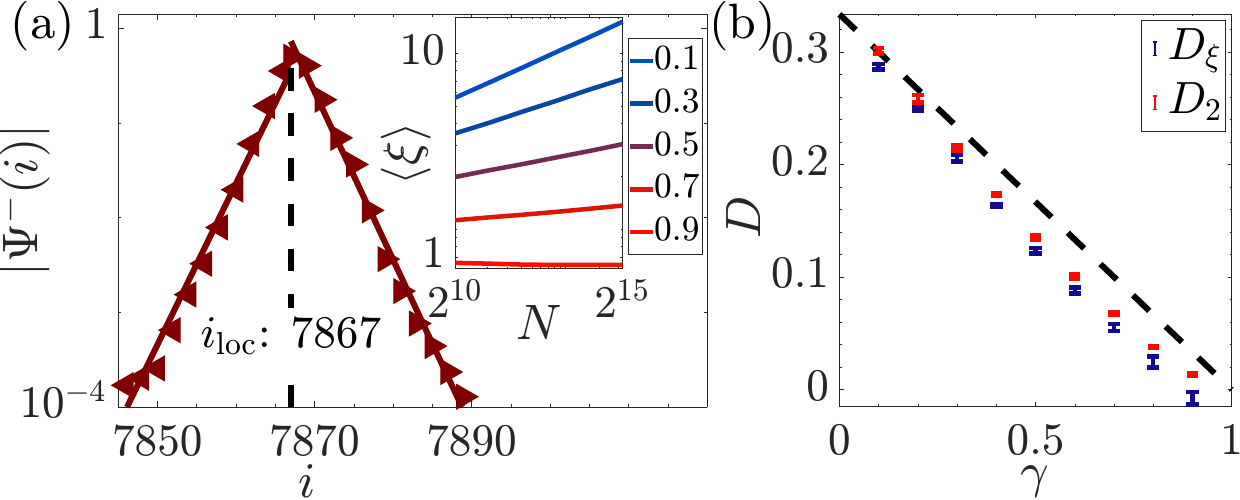}
	\caption{{\bf Ground-state exponential decay} (a)~Ground state components, $\Psi^-(i)$, vs.~lattice index, $i$, from a single realization of \bte\ for $N = 8192, \gamma = 0.5$ where $\iloc$ is the localization center, i.e., the location of the wave-function intensity maximum. Markers denote raw data and solid lines denote linear fit of $\ln|\Psi^-(i)|$ vs.~$|i - \iloc|$, Eq.~\eqref{eq_exp_decay}.
Inset shows the ensemble averaged decay length scale, $\xi$ vs.~the system size, $N$ for various $\gamma$.
		(b)~fractal dimensions of ground state vs.~$\gamma$ where the dashed line denotes $D = \frac{1-\gamma}{3}$. $D_2$ and $D_\xi$ extracted from the system-size scaling of the typical IPR, Eq.~\eqref{eq_IPRq}, and $\xi$, i.e.~$\mean{\ln \ipr_2}\propto -D_2\ln N$ and $\mean{\ln \xi}\propto D_\xi\ln N$ for $2^{10}\leq N\leq 2^{15}$.
Error-bars denote 95\% confidence interval.
	}
	\label{fig_1}
\end{figure}
\subsection{Ground-state fractal dimension}\label{subsec_GS_Dq}
In Fig.~\ref{fig_1}(a), we plot the absolute values of the ground state components w.r.t.~the lattice indices from a single realization of the \bte\ for $N = 8192$ and $\gamma = 0.5$. Even for a single realization, the exponential decay of the ground state, $\ket{\Psi^-}$ is immediately apparent, i.e.
\begin{gather}\label{eq_exp_decay}
|\Psi^-(i)| \sim \exp\del{-\dfrac{|i - \iloc|}{\xi}} \ ,
\end{gather}
where $\xi$ is the decay length scale and $\iloc$ is the localization center, where the eigenstate has the largest intensity. In the inset of Fig.~\ref{fig_1}(a), we show that the ensemble averaged decay length scale, $\mean{\xi}$ has a power-law scaling w.r.t.~to the system size, i.e.~$\mean{\xi}\propto N^{D_\xi}$. In Fig.~\ref{fig_1}(b), we show that $D_\xi \approx \frac{1-\gamma}{3}$ for $0\leq\gamma\leq1$. We find that the fractal exponent $D_2$, defined by the system-size scaling of the IPR for the ground state $\ket{\Psi^-}$, Eq.~\eqref{eq_IPRq}, is consistent with $D_\xi$, i.e., $D_2 \approx \frac{1-\gamma}{3}$,
We further observe that the anti-ground state, $\ket{\Psi^+}$, i.e. the eigenstate with the largest energy also has the same fractal dimension $\frac{1-\gamma}{3}$.

Note that, the typical bulk eigenstates have the fractal dimension $D_\mathrm{bulk} \sim 1-\gamma$, i.e.~three times that of the spectral edge states.
Moreover, in the Appendix we compare the the distribution of IPR of the bulk and ground states, see Fig.~\ref{fig_A1}.
We show that, unlike the distribution of IPR of the bulk states, which is bimodal and fat-tailed, see Fig.~\ref{fig_A1}(a), the corresponding ground-state IPR distribution is well described by a Gaussian distribution with both mean and standard deviation scaling as $N^{\frac{1-\gamma}{3}}$, see Fig.~\ref{fig_A1}(b).

\subsection{Localization center}\label{subsec_GS_iloc}
\begin{figure}[t]
	\centering
	\includegraphics[width=0.97\columnwidth]{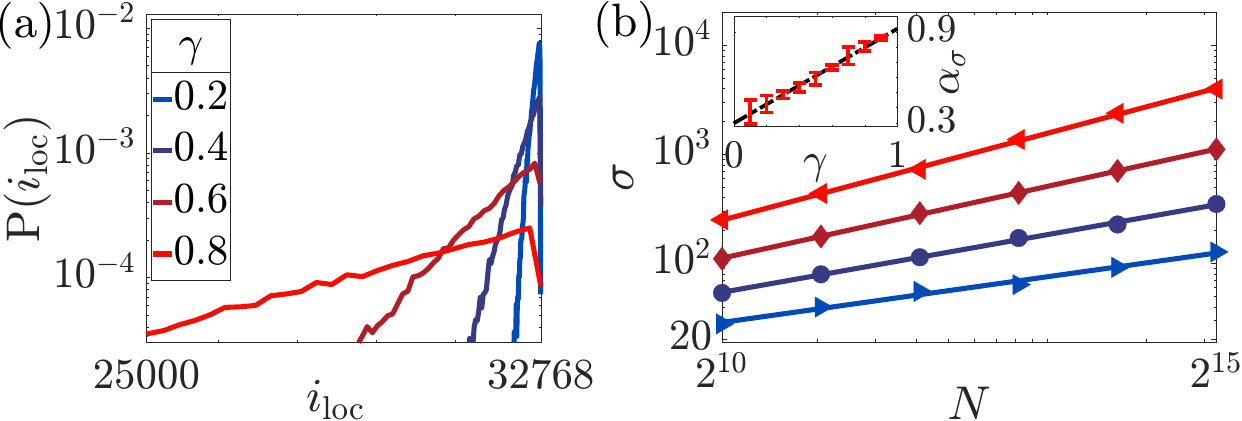}
	\caption{{\bf Ground-state localization center distribution} (a)~Probability distribution $\prob{\iloc}$ of the localization center, $\iloc$, of the ground states for various $\gamma$ and $N = 32768$.
		(b)~Width of $\prob{\iloc}$ vs.~$N$ for various $\gamma$, extracted from an exponential tail,~$\prob{\iloc}\propto \exp\del{\frac{\iloc}{\sigma}}$. The markers and solid lines denote raw data and linear fit in log-log scale, respectively. The inset shows the system-size scaling $\alpha_\sigma$ of $\sigma\propto N^{\alpha_\sigma}$ as a function of $\gamma$, where the dashed line denotes $\alpha_\sigma = \frac{2\gamma+1}{3}$.
	}
	\label{fig_2}
\end{figure}

Next, we consider the spatial distribution of the localization centers, $\iloc$, of the ground state. We find that the typical $\iloc$ scales linearly with $N$, Fig.~\ref{fig_A1}(c), i.e.~the ground state has the largest probability to stay near the right edge of the lattice. In Fig.~\ref{fig_2}(a), we show the spatial distribution of $\iloc$ for $N = 32768$ and various values of $\gamma$. The exponential tails of such distributions imply that $\prob{\iloc}\propto \exp\del{\frac{\iloc}{\sigma}}$, where $\sigma$ is the width of the distribution. Fig.~\ref{fig_2}(b) demonstrates that $\sigma$ increases with $N$ in a power-law manner while the system size scaling exponent, $\alpha_\sigma$, shown in the inset indicates that $\sigma \propto N^{\frac{2\gamma + 1}{3}}$.
The same scaling behaviors are also observed in case of the anti-ground state due to (i) the statistical homogeneity of the hopping terms $H_{nn}$, i.e.~$P(H_{nn})=P(-H_{nn})$ and (ii) the transformation $E\to -E$ leading to $\Psi_E(n) \to (-1)^n \Psi_{-E}(n)$ in Eq.~\eqref{eq_H_def}.

\subsection{Overlap of ground and anti-ground states $\ket{\Psi^\pm}$}\label{subsec_GS_overlap}
\begin{figure}[t]
	\centering
	\includegraphics[width=0.97\columnwidth]{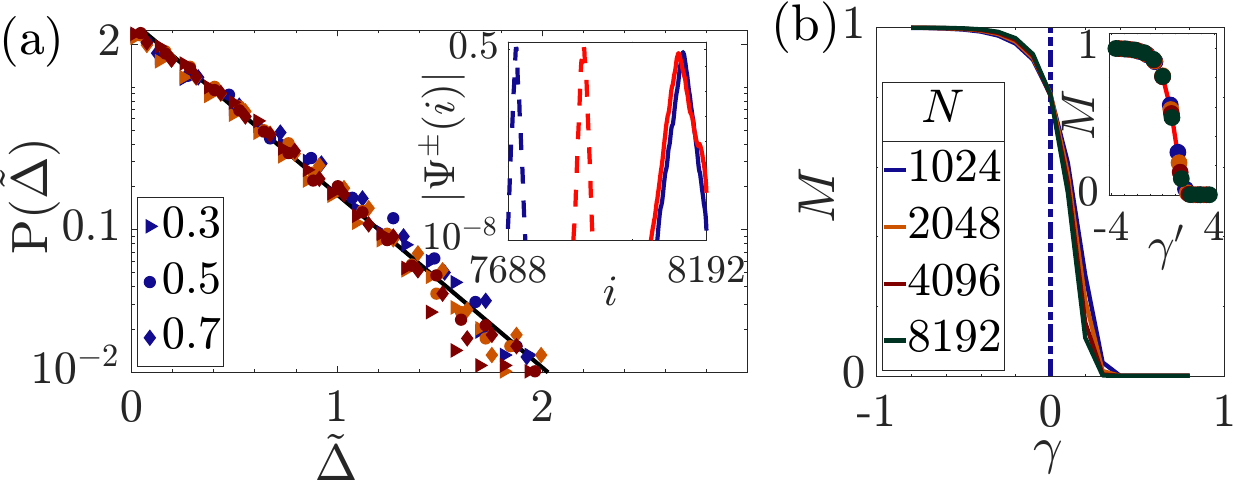}
	\caption{{\bf Correlations of ground and anti-ground states} (a) Probability distribution of the gap $\tilde{\Delta} = |\iloc^+ - \iloc^-|/\sigma$ of localization centers of the ground and anti-ground states, rescaled by $\sigma = \frac{3}{2\gamma+1}N^{\frac{2\gamma + 1}{3}}$. Different colors denote three different system sizes, $N = 1024, 4096, 16384$. Inset shows the ground (blue) and anti-ground (red) states from a single realization of \bte\ for $N = 8192$ where $\gamma = 0.3$ (solid) and $0.7$ (dashed).
		(b) Covariance of ground and anti-ground state vs.~$\gamma$ for various $N$. Inset shows collapsed data 
		where $\gamma' = \gamma\del{\log N}^\frac{1}{\nu}$ with $\gamma = 0$ being the critical point with the critical exponent $\nu\approx 1$~\cite{Das2}.
	}
	\label{fig_3}
\end{figure}

It is now pertinent to ask whether the ground and anti-ground states are correlated with each other, i.e., whether they are formed via a single strong level resonance or via an extensive set of many parametrically smaller ones.
To start with, in the inset of Fig.~\ref{fig_3}(a), we show the locations of the ground and the anti-ground states, $\ket{\Psi^\pm}$ from a single realization of the \bte\ for $N = 8192$ and $\gamma = 0.3$ and 0.7. One can see that $\ket{\Psi^\pm}$ has a significant degree of overlap close to the ergodic transition point, $\gamma = 0$, while  near the Anderson transition point, $\gamma = 1$, the overlap is almost negligible. To understand such an overlap with respect to various disorder realizations, we look at the random variable $\abs{\iloc^+ - \iloc^-}$, being the distance between the localization centers of the ground and anti-ground states. As shown in Fig.~\ref{fig_3}(a), the probability distribution of $\abs{\iloc^+ - \iloc^-}$ decays exponentially and has a width, $\sigma\propto \frac{3}{2\gamma + 1} N^{\frac{2\gamma + 1}{3}}$.
The distribution of the rescaled gap, $\tilde{\Delta}\equiv \frac{\abs{\iloc^+ - \iloc^-}}{\sigma}$, shown in Fig.~\ref{fig_3}(a), collapses onto a single curve for different system sizes and parameter values.
The same scaling of the widths, $\sigma$ of the distributions $\prob{\iloc^\pm}$ and $\prob{\abs{\iloc^+ - \iloc^-}}$ indicates that the localization centers of the ground and anti-ground states are independent from each other and identically distributed according to the exponential distribution $\propto \exp\del{N^{ -\frac{2\gamma + 1}{3}} \iloc^\pm }$.

To quantify the overlap of $\ket{\Psi^\pm}$ more accurately, we compute the covariance, $M = \sum_{i} |\Psi^-(i)\Psi^+(i)|$~\cite{Das5}. For two states with the complete overlap, $M = 1$ whereas $M\to 0$ for states with no significant hybridization. Fig.~\ref{fig_3}(b) shows that the covariance vs.~$\gamma$ for various system sizes show a crossover from $M = 1$ deep in the regime $\gamma<0$ to $M\to 0$ in the NEE phase ($0<\gamma<1$). Moreover, the crossover curves tend to intersect at $\gamma = 0$ and get steeper upon increasing $N$. The finite-size collapse of the data, shown in the inset of Fig.~\ref{fig_3}(b), suggests a $2$nd-order phase transition with a critical point at $\gamma=0$. For all values of $\gamma>0$, ground and anti-ground states do not hybridize in the thermodynamic limit, thus confirming the multiple-resonance nature of these states.

\subsection{Analytical consideration} \label{subsec_GS_analytics}
We now present an analytical understanding of the above numerical results.
Let us begin by recalling  that the $1$d Anderson model with constant hopping strength $t$ and disordered on-site potentials $\epsilon_n$ can be represented as a $2$d classical Hamiltonian map~\cite{Izrailev4, Izrailev5}:
\begin{align}
	\label{eq_H_map}
	\begin{split}
		\hat{H} =& \sum_{n = 1}^N \epsilon_n c_n^\dagger c_n + t \sum_{k = 1}^{N-1} c_k c_{k+1}^\dagger + c_k^\dagger c_{k+1}\\
		\lra& \begin{cases}
			x_{n+1} = x_n \cos\mu - (p_n + A_n x_n) \sin\mu\\
			p_{n+1} = x_n \sin\mu + (p_n + A_n x_n) \cos\mu
		\end{cases}
	\end{split}
\end{align}
where $c_k^\dagger, c_k$ are the creation and annihilation operators at the $k$th site and $(x_n, p_n)$ are the position and momentum of a linear kicked oscillator. In terms of the parameter $\mu$, we can express  the energy, $E = 2 t\cos\mu$ and the kick strength, $A_n = -\dfrac{\epsilon_n}{t \sin \mu}$.
In~\cite{Izrailev5}, it has been shown that in the weak disorder limit, $|A_n|\ll 1$, the localization length of the ground/anti-ground state is given by
\begin{gather}\label{eq_xi}
	\xi \propto \lrp{\frac{t^2}{\mean{\epsilon_n^2}}}^{{1}/{3}} \ . 
\end{gather}

On the other hand, it has been recently shown that for the \bte\ in the NEE phase, there exists a spatially local mapping to the 1d Anderson model with $N$-dependent hopping strengths~\cite{Das5}.
Specifically, a $1$d lattice of length $N$ governed by the \bte\ can be partitioned into nearly independent spatial blocks $\Delta_0, \Delta_1,\dots, \Delta_{l_\mathrm{max}}$ where the zeroth block represents first $N^\gamma$ sites, $\Delta_0 = [1, N^\gamma]$, and the $l$th block for $l\geq 1$ has the length $\abs{\Delta_l}\sim N^{\gamma+\zeta_l}$ and is defined as
\begin{align}
	\label{eq_block}
	\Delta_l \equiv [N^{\gamma + \zeta_l}, c N^{\gamma + \zeta_l}],\quad \zeta_l = (l-1) \frac{\ln c}{\ln N}
\end{align}
where $l_\mathrm{max} = \del{1-\gamma}\tfrac{\ln N}{\ln c}
$ and $c\sim \mathcal{O}(1)$ is a constant. On one hand, within $\Delta_0$, the hopping terms are negligible compared to the typical on-site potential $\mathcal{O}(1)$, hence all the sites within are effectively disconnected from the rest of the lattice and host single-site-localized eigenstates. On the other hand, the model in $\Delta_l$ for $l\geq 1$ can be shown to be asymptotically equivalent to the $1$d Anderson model of length $\abs{\Delta_l}\sim N^{\gamma+\zeta_l}$ with uncorrelated diagonal disorder $\mathcal{O}(1)$ and nearly homogeneous hopping $t \simeq y_{\Delta_l}\sim N^{\zeta_l/2}$ growing with $l$. This mapping works in the non-ergodic phases at $\beta\lessapprox 1$.

According to the above mapping, eigenstates are exponentially decaying at a length scale $\xi_l\sim N^{\zeta_l}$, hence the block $\Delta_l$ consists of $N^\gamma$ sub-blocks of length $N^{\zeta_l}$. In addition, each of these sub-blocks has Gaussian density of states (DOS) with bandwidth $N^{\frac{\zeta_l}{2}}$ and, thus, the mean level spacing $\delta_l \sim N^{-\frac{\zeta_l}{2}}$.

As a result, the last spatial block, $\Delta_{l_\mathrm{max}}$ ($\zeta_{l_\mathrm{max}} = 1-\gamma$) is the largest one containing a finite fraction $\mathcal{O}(N)$ of states, that exponentially decay within the length scale $N^{1-\gamma}$ and has the largest energy bandwidth $N^{\frac{1-\gamma}{2}}$ among all possible spatial blocks. Consequently, the ground state of the $\Delta_{l_\mathrm{max}}$ block should coincide with the ground state of the the entire \bte.

In the $\Delta_{l_\mathrm{max}}$ block, $t^2 \sim N^{1-\gamma}\gg 1$ and $\mean{\epsilon_n^2}=1$, correspond to the weak disorder limit of $|A_n|\ll1$ in~\cite{Izrailev5}, hence the decay length scale of the edge states should scale as in Eq.~\eqref{eq_xi}, i.e. $\xi\sim N^\frac{1 - \gamma}{3}$. This explains the fractal dimension of the ground (and anti-ground) state of $\Delta_{l_\mathrm{max}}$ and, thus, of the \bte\ to be $\frac{1-\gamma}{3}$, in contrast to the typical bulk states have a fractal dimension $1-\gamma$.
This supports the numerical results of Fig.~\ref{fig_1}.
Note that even at the ergodic transition, $\gamma=0$, both the ground and anti-ground states remain non-ergodic, with the fractal dimension $D_2 = 1/3$, while for the typical bulk states $D_2$ reaches its ergodic value $1$.
This is fully consistent with the results of the entire extended phase $\gamma<0$ in the next section and Fig.~\ref{fig_4}(c), where formally the above mapping is not applicable.

In order to understand the width of the distribution of the ground-state localization centers, one has to consider the inhomogeneity in the hopping terms within $\Delta_{l_\mathrm{max}}$.
As we have explained above, at the edge of the spectrum, the energy $E\sim \mathcal{O}(N^{\frac{1-\gamma}{2}})$, being large compared to the on-site potential $\epsilon_n\sim \mathcal{O}(1)$, is determined solely by the hopping amplitude $y_n\sim \sqrt{n N^{-\gamma}}$.
Thus, the ground state with localization center at $\iloc = N-\sigma$ should have the energy $E \sim \sqrt{(N-\sigma) N^{-\gamma}}$ and
\begin{align}
	\label{eq_map_1}
	\cos \mu = \frac{E}{2 t} = \sqrt{1 - \frac{\sigma}{N}} \approx 1 - \frac{\sigma}{2N}.
\end{align}
For the ground state of $\Delta_{l_\mathrm{max}}$ block, $\mu \sim t^{-2/3} = N^{-(1-\gamma)/3}\to 0$ in the weak disorder limit~\cite{Izrailev5}.
On the other hand, Eq.~\eqref{eq_map_1} implies that for $\mu\to 0$
\begin{align}
	\label{eq_map_2}
	\sigma\approx 2N (1-\cos\mu)\approx N \mu^2 \sim N^{\frac{2\gamma + 1}{3}}.
\end{align}
Therefore, the ground state is likely to have a localization center close to the right edge of the lattice ($N-\iloc\ll N$) and its distribution should have a width scaling as $\sigma\sim N^{\frac{2\gamma + 1}{3}}$. This confirms the numerical results, observed in Figs.~\ref{fig_2}, \ref{fig_3}(a) and \ref{fig_A1}(c).

All the above results are valid as long as the conditions for (i)~the local spatial mapping of~\cite{Das5} and (ii)~the weak disorder limit of~\cite{Izrailev5} are satisfied.
The latter is applicable in the delocalized phases, $\gamma<1$. Otherwise, all the states, including the ground state, are localized and results can be formally extended to the localized phase. 
On the other hand, the former local mapping is valid for $\gamma>0$.
Thus, the uncorrelated nature of the ground and anti-ground states of the $1$d Anderson model together with local mapping of~\cite{Das5} confirms the results of Fig.~\ref{fig_3}(b) that the correlations are absent in all the non-ergodic phases, $\gamma>0$. The constraint of local spatial mapping to $1$d Anderson model breaks down as soon as $\beta\gtrsim 1$. Hence, the region $\gamma<0$ needs to be treated separately, as discussed in the next section.

\section{Extended phase for $\gamma<0$}\label{sec_Erg}
\begin{figure}[t]
	\centering
	\includegraphics[width=0.97\columnwidth]{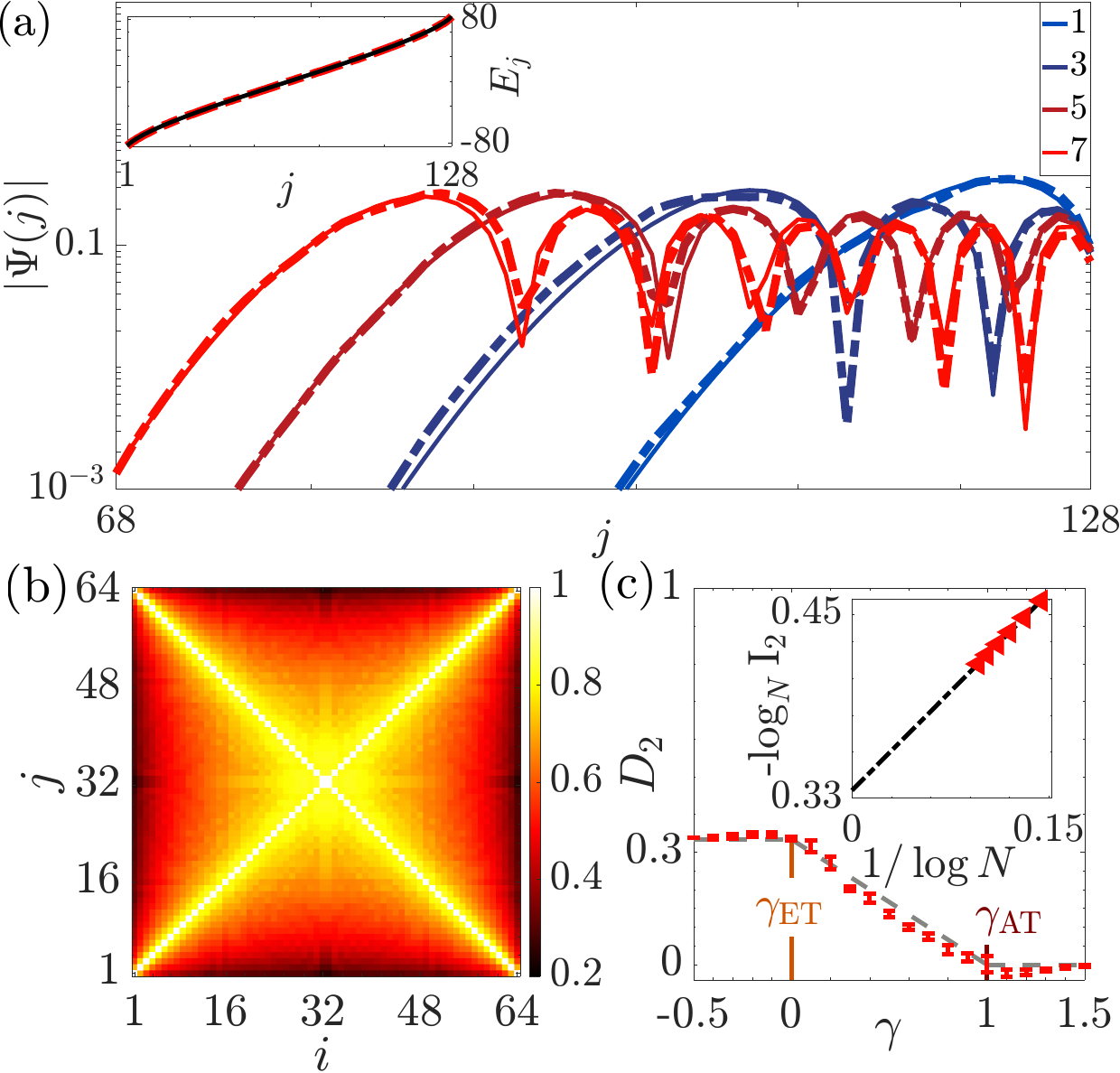}
	\caption{{\bf Eigenstate structure for $\gamma<0$} (a)~Ground and first few excited states from a single realization of the \bte\ for $N = 128$ and $\gamma = -0.7$ where in the legend, 1 (3) corresponds the ground (second excited) state and so on. Inset shows the eigenspectrum. In both main plot and the inset, dashed (solid) lines correspond to the exact-diagonalization numerical result (analytical expression in Eq.~\eqref{eq_H_mean_state}).
		(b)~Covariance $M_{ij}$ between all possible pairs of eigenstates $i$ and $j$ of $\Hm$ for $N = 64$.
		(c)~Fractal dimension $D_2$ for the ground state of \bte\ vs.~$\gamma$. Error-bars denote 95\% confidence interval. $\get=0$ and $\gat=1$ denote the ergodic and Anderson transitions, respectively. Inset shows finite-size fractal dimensions for $\Hm$ as a function of $1/\log N$, where extrapolation of $1/\log N\to 0$ (i.e.~$N\to\infty$) estimates $D_2 = 1/3$.
	}
	\label{fig_4}
\end{figure}

In the region $\gamma < 0$, the typical hopping amplitude, $y_n^\mathrm{typ} \sim \sqrt{ \frac{n N^{|\gamma|}}{2} }$ is large for all $n$ at $N\gg 1$ and $y_{n+m}-y_{n}\sim m N^\frac{|\gamma|}{2}/n \gg 1$ even for $m = 1$. Thus, the smallest mean hopping term is large compared to both the typical on-site terms and off-diagonal fluctuations $\sim \mathcal{O}(1)$. Consequently, a Hamiltonian $H$ from the \bte\ for $\gamma<0$ rescaled as $H\to \sqrt{2}N^\frac{\gamma}{2}H$ can be approximated by its mean
\begin{align}
	\label{eq_H_mean}
	\Hm_{m, n} = \sqrt{m} \delta_{m, n-1} + \sqrt{m - 1}\delta_{m, n+1}.
\end{align}
$\Hm$ is equivalent to $\hat{a}^\dagger + \hat{a}$ represented in the first $N$ eigenstates of the Harmonic oscillator where $\hat{a}, \hat{a}^\dagger$ are the corresponding annihilation and creation operators, respectively. Solving the characteristics equation $\det (\Hm - E\mathbb{I}) = 0$, we get the eigenstates of $\Hm$ as
\begin{align}
	\label{eq_H_mean_state}
	\Psi_{\lambda_n} (k) = \mathcal{Z}_n \frac{\mathcal{H}_{k-1}(\lambda_n)}{\sqrt{ 2^{k-1} (k-1)! }}
\end{align}
where $\mathcal{Z}_n$ is a normalization factor and $\Psi_{\lambda_n}(k)$ is the $k$th component of the eigenstate $\ket{\Psi_{\lambda_n}}$ with energy $\lambda_n$ which is the $n$th zero of the $N$th order Hermite polynomial
\begin{align}
	\label{eq_Hermite}
	\mathcal{H}_N(x) = (-1)^N e^{x^2} \frac{d^N}{dx^N} e^{-x^2}.
\end{align}
The energy of the original Hamiltonian $H$ from \bte\ is $E_n \approx N^{\frac{|\gamma|}{2}} \lambda_n$. For a single realization of the \bte\ with $N = 128$ and $\gamma = -0.7$, we show the ground state and a few excited states in Fig.~\ref{fig_4}(a) while the eigenvalues are shown in the inset. In the same figure, we also show the analytical expressions from Eq.~\eqref{eq_H_mean_state} valid for the Hamiltonian $\Hm$. We observe that the analytical estimates from Eq.~\eqref{eq_H_mean_state} match perfectly the results of the exact diagonalization in case of both energy levels and eigenstates, even for a single realization of \bte\ with a relatively small system size. Thus, $\Hm$ is indeed a good approximation of the \bte\ for $\gamma<0$ where exponential decay of the eigenstates observed in the NEE phase is absent.

The largest zero of the Hermite polynomial in Eq.~\eqref{eq_Hermite} is~\cite{Szego1}
\begin{align}
	\label{eq_Hermite_zero}
	\lambda^\mathrm{max} = \sqrt{2N} + \mathcal{O}(N^{-{1}/{6}}).
\end{align}
Thus, energy bandwidth of $\Hm$ is $\sim \sqrt{N}$. Then, we can estimate the bandwidth of original Hamiltonian $H$ as $\sqrt{N}\cdot N^\frac{|\gamma|}{2} = N^\frac{1-\gamma}{2}$ and the mean level spacing as $\delta\sim N^{-\frac{\gamma+1}{2}}$. Both of these scalings match with those of the \bte\ for $\gamma<0$.

The eigenstates in Eq.~\eqref{eq_H_mean_state} along with the symmetry $\mathcal{H}_n(-x) = (-1)^n\mathcal{H}_n(x)$ are such that the absolute values of the components of two eigenstates with energy $\pm E$ are equal. Hence, the $j$th and $(N+1-j)$th energy-ordered states should fully hybridize with each other and show perfect correlations as shown in Fig.~\ref{fig_4}(b), inferred from the covariance of each pair of eigenstates. In particular, the ground state, $\ket{\Psi^-}$ and the anti-ground state, $\ket{\Psi^+}$ have perfect overlap in the case of $\Hm$. As $\Hm$ is a good approximation of \bte\ for $\gamma<0$, $\ket{\Psi^\pm}$ of \bte\ should have high degree of overlap which is reflected by $M$ being close to unity in Fig.~\ref{fig_3}(b) for $\gamma < 0$, where $M$ is the covariance of $\ket{\Psi^\pm}$ quantifying their hybridization.

Additionally in Appendix~\ref{sec_C_r}, we show that the level-spacing ratio, $r$-statistics~\cite{Oganesyan2007,Atas1} at the spectral edges has a value $\mean{r}\simeq 0.8187$ for $\Hm$. Such a value of $r$-statistics is obtained in case of \bte\ for $\gamma<0$ upon increasing system size or decreasing $\gamma$ as shown in Fig.~\ref{fig_A2}(a). In contrast, $\mean{r}\to 1$ in the bulk spectrum as we get a rigid picket-fence spectrum for $\gamma<0$ and $N\gg 1$.

\subsection{Ground state for $\gamma<0$}
Finally to explain the behavior of low-energy eigenstates as in Fig.~\ref{fig_4}(a), we need to look at the expansion of the Hermite polynomials for large argument
\begin{align}
	\label{eq_Hermite_expansion_large}
	\mathcal{H}_n(x)_{|x|\to \infty} = (2x)^n\del{1 - \frac{n(n-1)}{(2x)^2} + \mathcal{O}(x^{-4})}.
\end{align}
The 2nd term in above expansion can be neglected for $x \gtrsim \dfrac{\sqrt{n(n-1)}}{2}$ or equivalently at $n\lesssim 2x$. Then, the scaling of $E$ with $N$ in $x$, Eq.~\eqref{eq_H_mean_state}, implies that for $|E|\gtrsim N^{-\frac{\gamma}{2}}\sqrt{\dfrac{n(n-1)}{2}}$, or equivalently for $n\lesssim |E|N^{\frac{\gamma}{2}}$, $\mathcal{H}_n(\lambda) \sim (|E|N^\frac{\gamma}{2})^n$ and, thus, the eigenstate with energy $E$ decays with $m = N-n$ as
\begin{align}
	\label{eq_H_mean_state_decay}
	\Psi_E(m) = \mathcal{Z} \frac{(|E|N^{\frac{\gamma}{2}})^{N-m}}{\sqrt{(N-m)!}} \approx \mathcal{Z} \del{\frac{|E|}{N^\frac{1-\gamma}{2}} \sqrt{ \frac{e}{1 - \frac{m}{N}} } }^{N-m}.
\end{align}
The largest root of the $n$th Hermite polynomial $\mathcal{H}_n(x)$ is at $x \approx \sqrt{2n}$, see Eq.~\eqref{eq_Hermite_zero} for $N\to n$. Then, $\mathcal{H}_n(x)$ oscillates as a function of $n$ without causing any decay in the eigenstate~\eqref{eq_H_mean_state} at $n\gtrsim \dfrac{x^2}{2} = \dfrac{E^2}{3N^{-\gamma}}$.
On the other hand, there is a decay slower than that in Eq.~\eqref{eq_H_mean_state_decay} within the interval $2x\lesssim n\lesssim\frac{x^2}{2}$, where no zeroes of $H_n(x)$ are present. All these three intervals describe the spatial behavior of the eigenstates for $\gamma<0$, see Fig.~\ref{fig_4}.

In the inset of Fig.~\ref{fig_4}(c), we show the finite-size fractal dimension $D_2^{(N)}\equiv -\log_N \ipr_2$ of $\Hm$ as a function of $1/\log N$, indicating that $D_2\approx D_2^{(N)} - c/\log N$ where $D_2 \approx 1/3$ is the true fractal dimension of the ground state of $\Hm$. In Fig.~\ref{fig_4}(c), we show $D_2$ as a function of $\gamma$ for the \bte. We find that $D_2 \approx 1/3$ for $\gamma\leq 0$, therefore matching with the prediction from $\Hm$. Therefore, the ground state has a fractal dimension three times smaller than that of the bulk states in the entire parameter regime of the \bte.

\section{Conclusions}\label{sec_concl}
Motivated by the peculiarity of the ground-state physics in quantum chaotic many-body systems~\cite{AtasBogomolny2012_MF_GS}, we consider the ground and anti-ground states of the \bte\ in this work.
We show that both in the extended phase ($\gamma<1$), the ground state of \bte\ exhibit non-ergodic properties with the fractal dimension, being three times smaller than its bulk value. The localization centers of these spectral edge states are exponentially distributed close to the most delocalized states, with the width, scaling non-trivially with the system size.

The main difference between the low-energy physics of the region $\gamma<0$ and non-ergodic phase ($0<\gamma<1$) is the correlations between the ground and the anti-ground states.
In the non-ergodic phase, these two edge states do not hybridize while the respective localization centers and the energies are uncorrelated.
In contrast, the correlations become rigid for $\gamma<0$: adjacent level repulsion is strong while the ground and anti-ground states have a perfect overlap.

In both the localized and the extended non-ergodic phases, the spatial local mapping to the $1$d Anderson model as suggested in~\cite{Das5}, provide analytical understanding of the above results in the NEE phase.
Using the standard weak-disorder limit for the spectral edge states, we found the ground-state fractal dimension to be one-third of its bulk value while the ground and anti-ground states remain uncorrelated.
On the other hand for $\gamma<0$, the physics is controlled by the mean Hamiltonian matrix and we have explicitly computed the corresponding eigenstates and eigenvalues. Hence, we claim that similar to the quantum chaotic many-body models, \bte\ shows a peculiar fractal structure of the ground-state eigenfunctions, even for highly correlated energy spectrum ($\gamma<0$). It will be of particular interest to see how the non-ergodicity of the ground state affects the finer structures of the bulk states, e.g.~the possibility of Hilbert space blockade~\cite{Haque1, Huang_NPB2019_universal_EE, Huang_2021_universal_EE, Pausch2021chaos,Pausch2021chaos2,Vidmar2023_Sent,sarkar2023entanglement}. Such a constraint can enforce weak ergodicity leading to non-trivial thermalization, similarly to other random-matrix models~\cite{Khaymovich2,Monthus,Das6} and random graphs~\cite{Biroli2017delocalized,Bera2018return,DeTomasi2019subdiffusion,Biroli2020anomalous,colmenarez2021SFF}.

\begin{acknowledgements}
	A.~K.~D. is supported by an INSPIRE Fellowship, DST, India and the Fulbright-Nehru grant no.~2879/FNDR/2023-2024.
I.~M.~K. acknowledges the support
by the European Research Council under the European
Union's Seventh Framework Program Synergy ERC-2018-SyG HERO-810451.
\end{acknowledgements}

\appendix
\renewcommand\thefigure{\thesection.\Roman{figure}}
\setcounter{figure}{0}
\renewcommand\thetable{\thesection.\Roman{table}}
\setcounter{table}{0}

\section{Energy-window size of spectral edge physics in ergodic phase}\label{sec_B_bound_edge}
\begin{figure}[t]
	\centering
	\includegraphics[width=0.97\columnwidth]{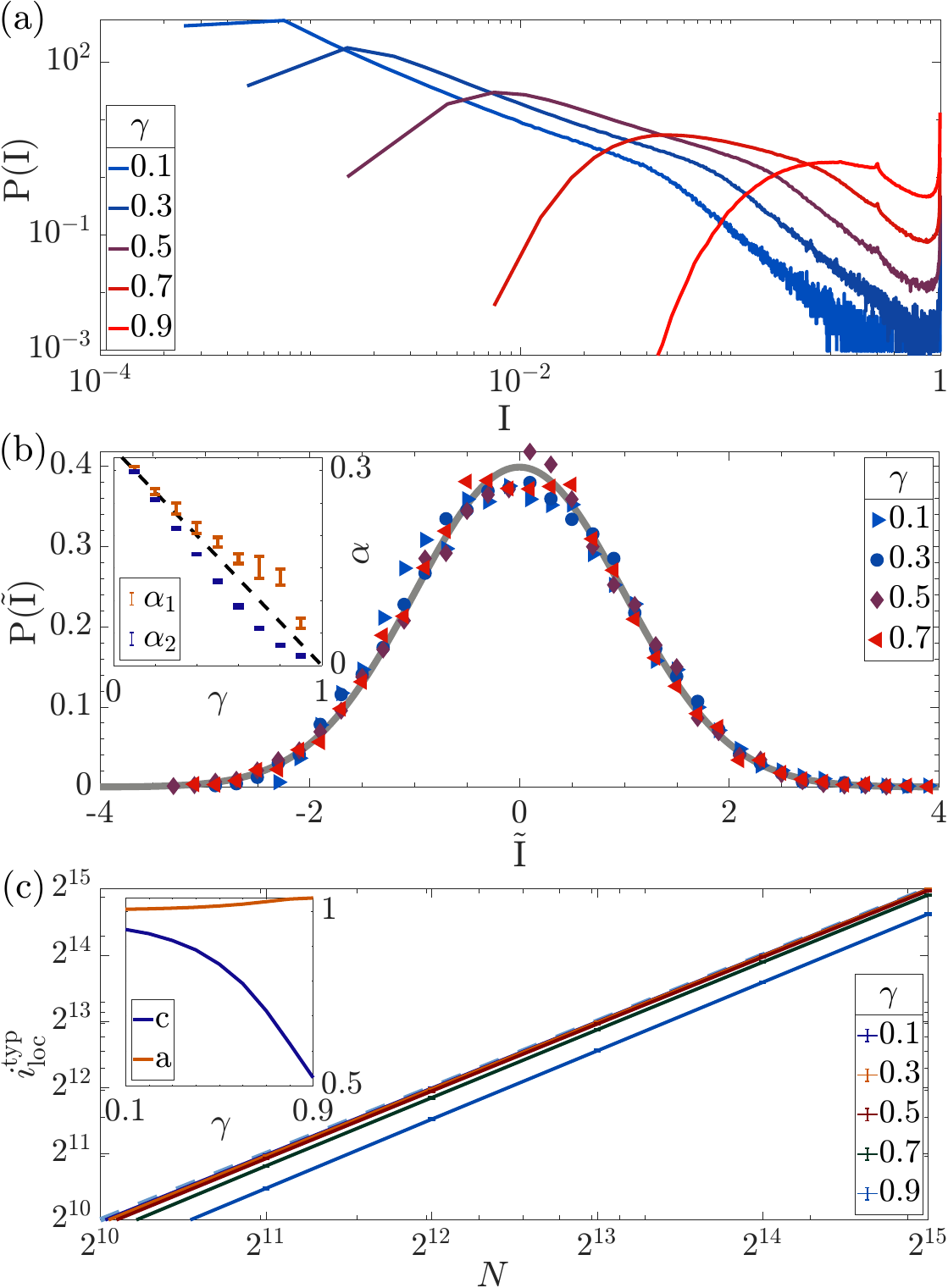}
	\caption{{\bf Comparison of bulk- and ground-state IPR distributions.}
(a)~Probability distribution of IPR of all the eigenstates in log-log scale for various $\gamma$ and $N = 8192$.
		(b)~The collapse of the probability distribution of $\tilde{\ipr}$ of ground state for $N = 8192$, where $\tilde{\ipr} = \frac{\ipr - \mean{\ipr}}{SD(\ipr)}$ stands for the rescaled fluctuations, relative to the standard deviation (SD), while the solid line denotes normal distribution, $\mathcal{N}(0, 1)$. Inset shows SD$\del{\ipr}\propto N^{-\alpha_1}$, $|\Psi^-(\iloc)|^2\propto N^{-\alpha_2}$ where dashed line denotes $\alpha = D^- \equiv \frac{1-\gamma}{3}$.
		(c)~typical ground-state localization center $\iloc$ vs.~$N$ where dashed line denotes $\iloc^\mathrm{typ} = N$. Inset shows $c$ and $a$ vs.~$\gamma$ where $\iloc^\mathrm{typ} = cN^a$.
	}
	\label{fig_A1}
\end{figure}
In the parameter region $\gamma \leq 0$, the DOS of \bte\ follows the semi-circle law, $\rho(E) = \dfrac{2}{\pi}\sqrt{1 - E^2}$ upon scaling the energy spectrum as $E\to E/2\sqrt{\mean{E^2}}$ where $\mean{E^2}\approx N^{1-\gamma}$ is the variance of the DOS~\cite{Das5}. In the interval $[1-\Delta, 1]$ if only the edge states are present, then $N\int_{1-\Delta}^{1} dE\rho(E) = \mathcal{O}(1)$. Note that~\cite{Majumdar2014}
\begin{align}
	\label{eq_edge_bound}
	\begin{split}
		\int_{1-\Delta}^{1} dE\rho(E) &= \frac{1}{\pi} \del{2\sin^{-1}\sqrt{\frac{\Delta}{2}} + (\Delta - 1)\sqrt{\Delta (2-\Delta)} }\\
		&= \frac{4\sqrt{2}}{3\pi}\Delta^\frac{3}{2} + \mathcal{O}(\Delta^\frac{5}{2})\\
		\Rightarrow N \times \Delta^\frac{3}{2} &= \mathcal{O}(1)\\
		\Rightarrow \Delta &= \mathcal{O}(N^{-\frac{2}{3}})
	\end{split}
\end{align}
Then, actual width of the energy window containing the edge states is
\begin{align}
	\label{eq_edge_bound_2}
	\Delta_\mathrm{edge} = \sqrt{\mean{E^2}}\times \Delta = \mathcal{O}(N^{-\frac{3\gamma + 1}{6}}).
\end{align}

\section{$r$-statistics at spectral edges in ergodic phase}\label{sec_C_r}
\begin{figure}[t]
	\centering
	\includegraphics[width=0.97\columnwidth]{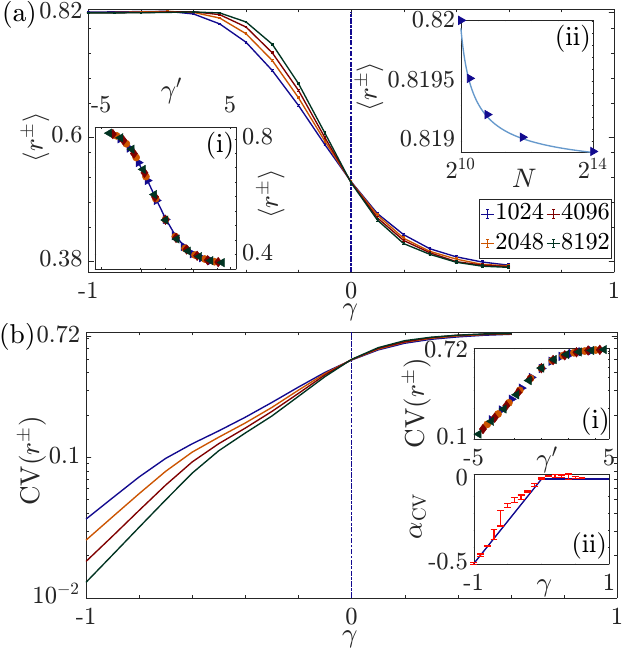}
	\caption{{\bf Level-spacing ratio $r^\pm$ at edge of the spectrum:} (a)~ensemble averaged $\mean{r^\pm}$. Inset~(i) shows the collapsed data assuming $2$nd order phase transition.
    Inset~(ii) shows $\mean{r^\pm}$ for the mean Hamiltonian vs.~$N$, Eq.~\eqref{eq_H_mean}. Solid line denotes power-law fitting: $\mean{r^\pm} = 0.1158 N^{-0.6543} + 0.8187$.
		(b)~Coefficient of variation, $\mathrm{CV}(r)\equiv\sqrt{\mean{r^2} - \mean{r}^2}/\mean{r}$. Inset (i) shows collapsed data whereas inset (ii) shows system size scaling, i.e.~CV$(r^\pm)\propto N^{\alpha_\mathrm{CV}}$.
	}
	\label{fig_A2}
\end{figure}
To quantify the short-range energy correlations at the edge of the spectrum, we look at the ratio of the level spacing between the edge state energies, $r^\pm = \min\cbr{\tilde{r}^\pm, \dfrac{1}{\tilde{r}^\pm}}$, where $\tilde{r}^- = \frac{E_3 - E_2}{E_2 - E_1}$ and $\tilde{r}^+ = \frac{E_{N} - E_{N-1}}{E_{N-1} - E_{N-2}}$~\cite{Oganesyan2007,Atas1}, assuming the eigenvalues are in the ascending order. In Fig.~\ref{fig_A2}(a), we show ensemble averaged $r^\pm$ as a function of $\gamma$ for various system sizes. Inset~(i) shows that $r^\pm$ undergoes a 2nd order transition at $\gamma = 0$. In inset~(ii), we show the $r^\pm$ for the Hamiltonian $\Hm$ in Eq.~\eqref{eq_H_mean} as a function of system size, which can be fitted using a power-law function. Thus, $\mean{r^\pm}\to 0.8187$ in the thermodynamic limit for $\Hm$, which is also the $r^\pm$ observed for the \bte\ in the ergodic regime for $N\gg 1$. Note that, in the bulk spectrum, the level-spacing ratio becomes unity in the thermodynamic limit owing to the full rigidity of the picket-fence structure of the energy spectrum \cite{Pandey6}.

In Fig.~\ref{fig_A2}(b), we show the coefficient of variation of $r^\pm$ as a function of $\gamma$. We observe that CV($r^\pm$) also undergoes 2nd order transition at $\gamma = 0$ (inset (i)). For $\gamma > 0$, CV($r^\pm$) converges towards the Poisson limit,
\begin{align}
	\label{eq_r_CV_psn}
	\mathrm{CV}_\mathrm{Poisson} = \sqrt{\frac{3 - 2\ln 4}{(\ln 4 - 1)^2} - 1} \approx 0.723855
\end{align}
Inset (ii) shows the system size scaling of CV$(r^\pm)\propto N^{\alpha_\mathrm{CV}}$ where
\begin{align}
	\label{eq_r_CV_scale}
	\alpha_\mathrm{CV} = \begin{cases}
		\frac{\gamma}{2}, & \gamma < 0\\
		0, & \gamma\geq 0
	\end{cases}
\end{align}


\end{document}